\documentclass[conf, onecolumn]{ao4elt3}
\usepackage{graphics}  
\usepackage[varg]{txfonts} 
\usepackage[latin1]{inputenc}
\usepackage{textgreek}
\begin{document}
\pagestyle{plain}
\title{LYOT-BASED ULTRA-FINE POINTING CONTROL SYSTEM FOR PHASE MASK CORONAGRAPHS}

\author{Garima Singh\inst{1,2}\thanks{singh@naoj.org} \and Frantz Martinache\inst{1} \and Pierre Baudoz\inst{2} \and Olivier Guyon\inst{1} \and Taro Matsuo\inst{3} \and Christophe Clergeon\inst{1,2}}
\institute{Institute name, Department, Address, Country \and Institute  name, Department, Address, Country \and ...}
\institute{National Astronomical Observatory of Japan, Subaru Telescope, 650 N A'Ohoku Pl, Hilo, HI  96720, USA \and LESIA, Observatoire de Paris-Meudon, 5 Place Jules Janssen, F-92195 Meudon Cedex, France \and Kyoto University, Kitashirakawa-Oiwakecho, Sakyo-ku, Kyoto 606-8502, Japan}
\abstract{
High performance coronagraphic imaging at small inner working angle requires efficient control of low order aberrations. The absence of accurate pointing control at small separation not only degrades coronagraph starlight rejection but also increases the risk of confusing planet's photons with starlight leaking next to the coronagraph focal plane mask center. Addressing this issue is essential for preventing coronagraphic leaks, and we have thus developed a new concept, the Lyot-based pointing control system (LPCS), to control pointing errors and other low order aberrations within a coronagraph. The LPCS uses residual starlight reflected by the Lyot stop at the pupil plane. Our simulation has demonstrated pointing errors measurement accuracy between 2-12 nm for tip-tilt at 1.6 {\textmu}m with a four quadrant phase mask coronagraph. 
}
\maketitle
\section{Introduction}
\label{intro}
The direct detection and characterization of extrasolar planets is affected by the rapidly changing atmosphere. The optical systems can not perform the high contrast imaging without the accurate measurement and calibration of the wavefront. But with the help of Adaptive Optics, the optical system are able to reach the diffraction-limit and the post processing techniques such as Differential Imaging has made it possible to identify faint companions at angular separation $\gtrsim$10 $\lambda$/D from their parent star (Marois et al. 2008 and Lagrange et al. 2009). 
Moreover, the newly developed small inner working angle (IWA) coronagraphs employed on Extreme Adaptive Optic Systems (ExAO) have improved their ability to image exoplanets within a few $\lambda$/D. But Guyon et al. (2006) addresses the following challenges that limits coronagraphic performance: stabilizing starlight at the center of the coronagraph; isolating companion's photons by calibrating speckles from star; obtaining high star-planet contrast ($10^{-6}$ in IR) over large spectral bandwidth.

The closer the search area is to the star, the more difficult it is to image the companions because high performance coronagraphs are extremely sensitive to tip/tilt errors (Lloyd $\&$ Sivaramakrishnan 2005; Shaklan $\&$ Green 2005; Sivaramakrishnan et al. 2005; Belikov et al. 2006; Guyon et al. 2006). Thus, their performance is limited by how well the low order wavefront aberrations are controlled and calibrated. 

The Subaru Telescope Coronagraphic Extreme AO (SCExAO) is a flexible, high performance coronagraphic system that can detect high contrast structures as close as 1$\lambda$/D and which uses a  coronagraphic low-order wavefront sensor (CLOWFS) as described in Guyon et al. (2009) and Vogt et al. (2011) to suppress low order aberrations in the coronagraph. The CLOWFS uses a dual-zone focal plane mask and analyzes the defocused image of the unused partially reflected starlight. The recent laboratory demonstration of CLOWFS on NASA's High Contrast Imaging Testbed (HCIT) at JPL has shown the stabilization of tip-tilt with 0.001$\lambda$/D residual in closed loop for $\lambda$ = 808 nm on Phase Induced Amplitude Apodization (PIAA)-type coronagraph (Kern et al. 2013).

To further enhance the IWA of SCExAO, we have modified the instrument to support phase mask coronagraphy with a goal IWA $<$ 1$\lambda$/D. To be able to take advantage of the small IWA offered by phase mask coronagraphs (PMCs), we introduce a new generation of CLOWFS to do the pointing control which is compatible with PMCs. It is based on the idea of re-imaging the starlight reflected by the Lyot Stop towards an imaging camera at a defocused position. We call our new system as Lyot-based pointing control system (LPCS). The common property of the PMCs is that they diffract starlight outside of the geometrical pupil which is then blocked by the conventional Lyot stop. For super fine controlling of the pointing errors, we have modified the Lyot design for them. We used a reflective Lyot stop (RLS) which fully reflects the unused starlight towards the low order wave front sensor measuring the low order modes accurately. 

The concept of RLS and pointing error estimation theory is presented in section 2. We describe the LPCS optical configuration, numerical simulation and their results in section 3. All of our simulations are done without considering photon noise.
 \begin{figure}[h]
   \centerline{
        \resizebox{0.7\columnwidth}{!}{\includegraphics{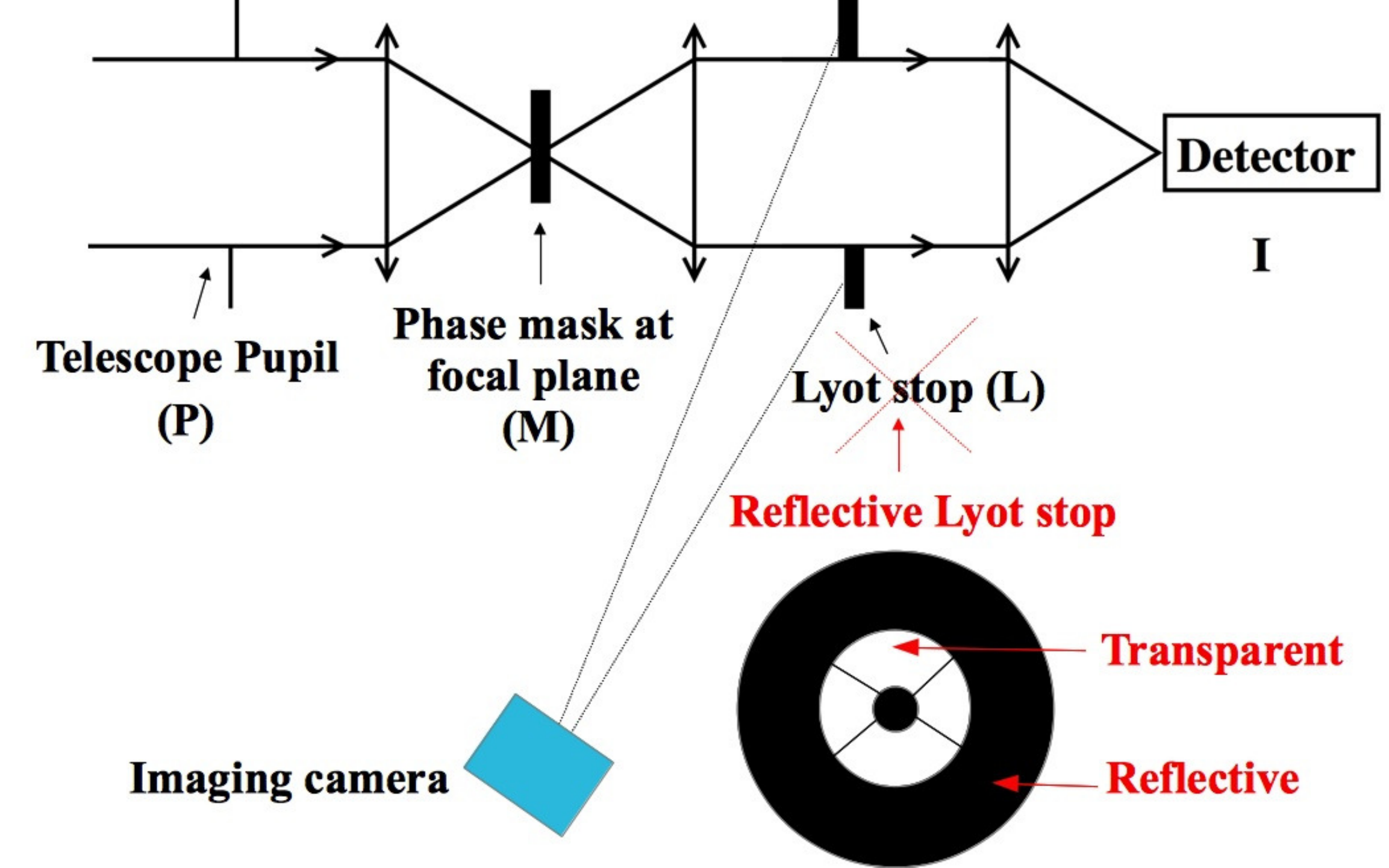}}}
   \caption{
   Basic design of the LPCS consisting of a high performance coronagraph combined with reflective Lyot stop whose geometry can be adopted considering telescope pupil with central obstruction and spider arms.  }
  \label{phase1}
\end{figure}
\section{Principle}
\subsection{A reflecting Lyot stop wavefront sensor}
Atmospheric turbulence and telescope pointing errors make it difficult to center the stellar light on occulting mask which creates halo of speckles around the occulter preventing detection of companion. SCExAO's CLOWFS has successfully used a focal plane mask as a means of measuring and correcting the low order aberrations and has made planet detection possible within the 1 - 2$\lambda$/D region.
Its possible to reach the region $\lesssim$ 1$\lambda$/D with high performance PMCs, but the sensitivity to tip-tilt errors make it mandatory to correct for low order aberrations to obtain the best starlight rejection. However a reflective focal plane mask as used in SCExAO is not feasible  to measure pointing errors for PMCs, hence a new solution is needed.\\
The PMCs have the tendency to redistribute the energy spatially in the telescope pupil, canceling on-axis light and diffracting it outside the aperture which is then absorbed by the Lyot stop. 
We introduce a new concept where we fully utilize the unused diffracted light by reflecting it via Lyot stop (called as reflective Lyot stop) towards an imaging camera to accurately measure pointing errors. Fig. \ref{phase1} describes the basic concept of LPCS. At the focal plane of the telescope, a phase mask diffracts starlight in a re-imaged pupil plane. This residual starlight is then collected by the RLS (L) and then reflected towards the camera which acquires a defocused starlight image. 
\subsection{Pointing errors estimation based on linearity approximation}
\label{LA}
\begin{figure}
   \centerline{
        \resizebox{0.8\columnwidth}{!}{\includegraphics{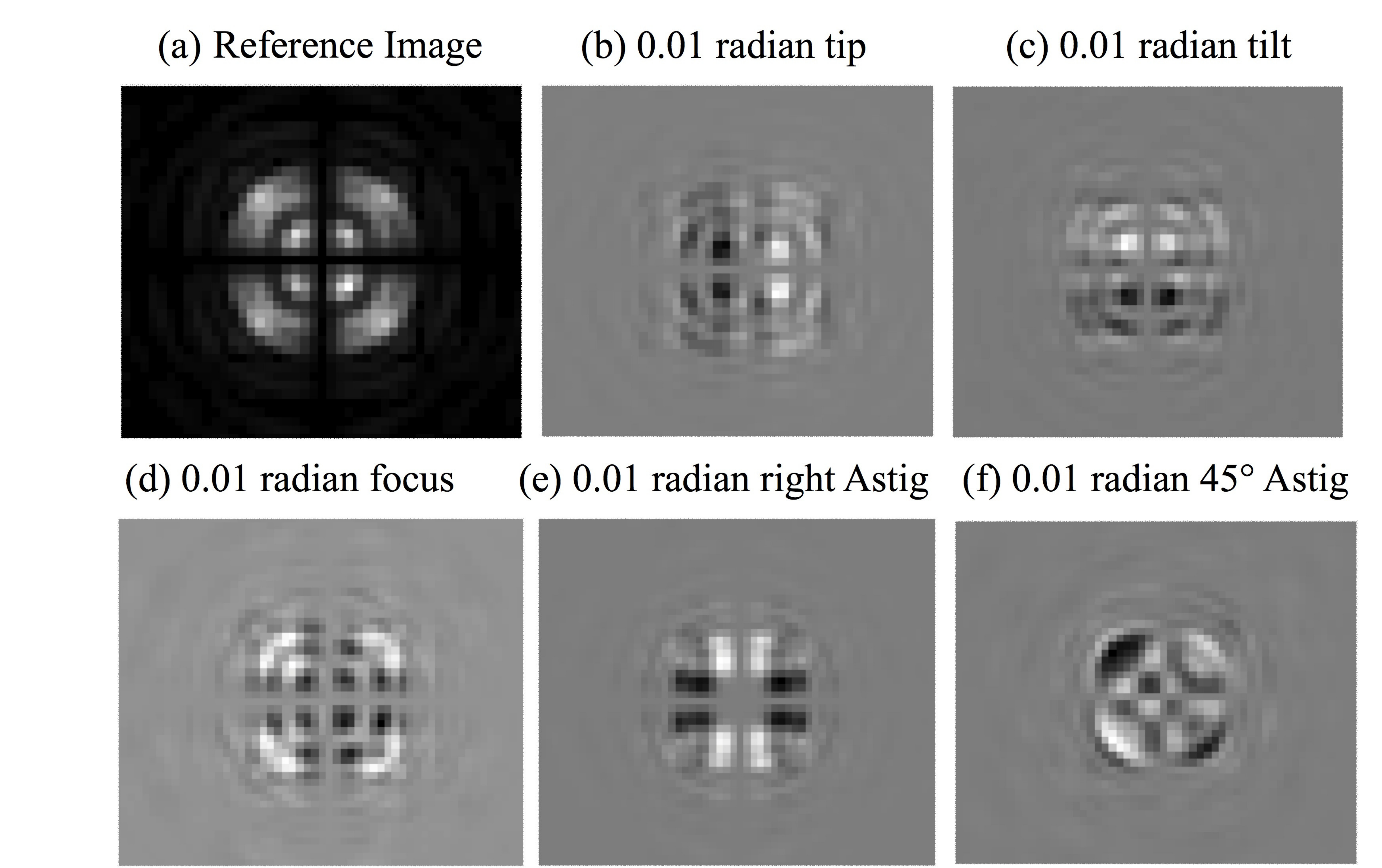}}}
   \caption{
 Response of the LPCS to the low-order modes which are our calibration frames to measure the unknown aberrations present in the entrance pupil.
   }
  \label{cal}
\end{figure}

In a post-AO correction scenario, residual phase errors can be assumed to be small ($<<$1 radian phase rms at pupil). Simulation work, as well as preliminary experimental results show that our Lyot-based pointing control system (LPCS) is a linear wavefront sensor. For small pointing errors, intensity fluctuations in the RLS image appear to be a linear function of low-order phase errors before the coronagraph. Calling $I_{0}$ our reference image, acquired by the LPCS camera with no tip-tilt, and $I_{R}$ the instant LPCS image, we can relate the difference between these two images to a linear combination of modes. If one considers tip-tilt alone, then we can for instance write:
\begin{equation}
I_{R(\alpha_{x},\alpha_{y})} - I_{0} =  \alpha_{x} S_{x} + \alpha_{y} S_{y}
\label{1}
\noindent
\end{equation}
where $S_{x}$ and $S_{y}$ represent the sensor's respective response to tip and tilt. For any instant image $I_{R}$, one can therefore identify unknown tip-tilt $(\alpha_x,\alpha_y)$, by direct projection on the basis of modes, or using a least square algorithm. $S_{x}$ and $S_{y}$ are obtained during a calibration step, after applying a controlled amount of tip and tilt modes. Example of response modes are shown in Fig. \ref{cal}, for tip-tilt as well other modes: defocus and astigmatism, showing that the LPCS is a versatile sensor.
\begin{figure}
   \centerline{
        \resizebox{0.8\columnwidth}{!}{\includegraphics{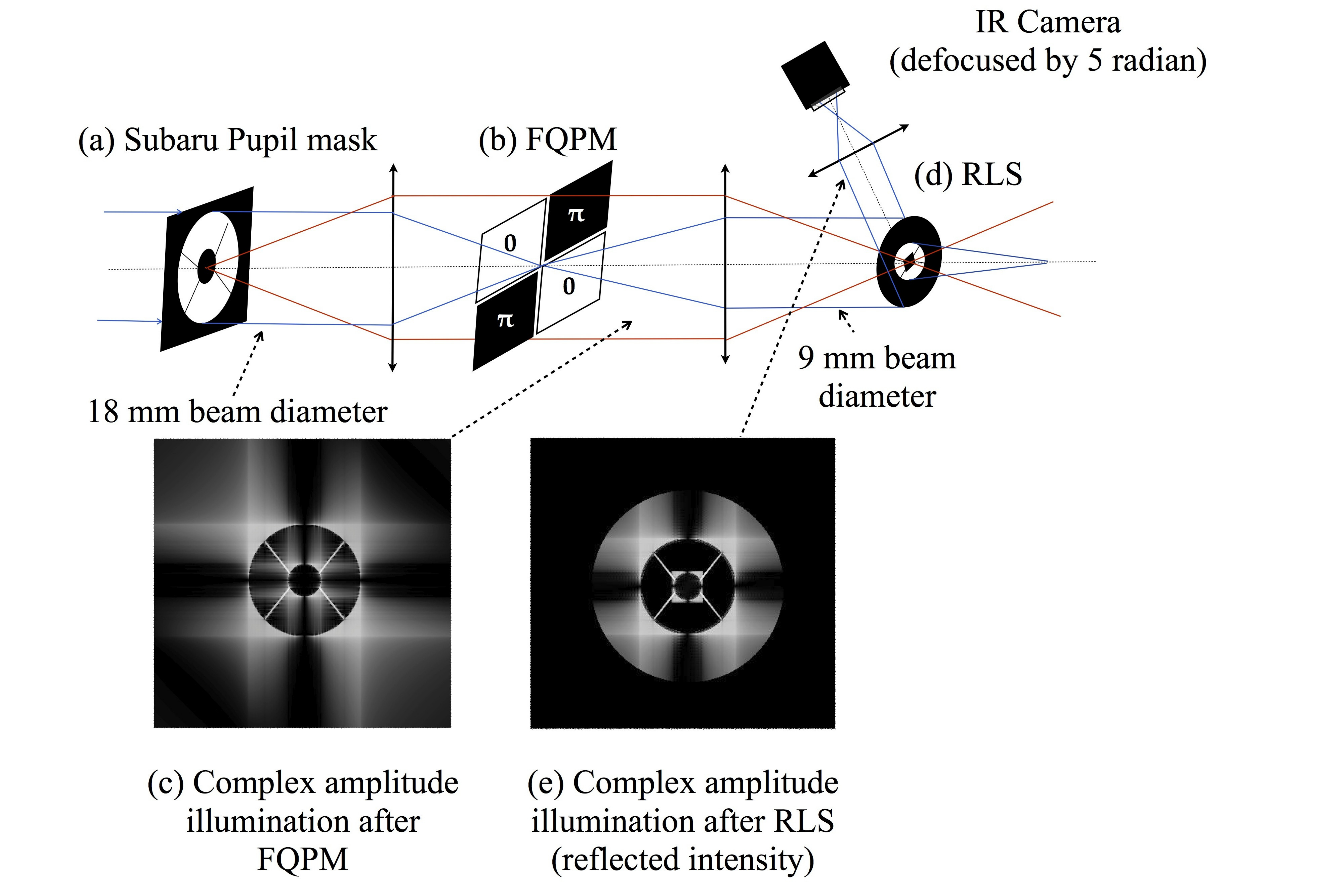}}}
   \caption{
  Simplified optical layout of revised SCExAO system showing LPCS with FQPM as an example. (a) Subaru pupil mask. (b) Four quadrant phase mask (FQPM). (c) Intensity pattern after FQPM. (d) Reflected Lyot Stop (RLS). (e) Complex amplitude as seen at RLS plane. See section \ref{OC} for more details.
   }
  \label{lpcs}
\end{figure}
\section{Lyot-based pointing control system}
\label{param}
\subsection{ Optical Configuration}
\label{OC}
We have developed a simulation tool by reproducing the revised SCExAO system (Jovanovic et al. 2013) at Subaru Telescope to test LPCS concept. A simplified optical configuration of the LPCS is shown in Fig. \ref{lpcs}. The optical components are described as follows:
\begin{itemize}
\item{Entrance Pupil $(a)$: $f14$ beam of 18 mm diameter as an input to SCExAO which is the output of 8-meter Subaru Telescope's Adaptive Optics AO188 system)}
\item{Phase Mask Coronagraph $(b)$: FQPM (Optimized wavelength $\lambda$=1.6 {\textmu}m).
The FQPM divides the focal plane in four quadrant and provides a $\pi$ phase shift in the two opposite quadrants, resulting in self-destructive interference in the relay pupil (Rouan et al. 2000). To be more realistic, we simulated a FQPM with manufacturing defects:  gap of 2.5 {\textmu}m in the transition zone of the opposite quadrants and the shift of $\approx$1.8 {\textmu}m between the layers deposited on the two adjacent quadrants.}
\item{Reflective Lyot stop $(d)$: $f28$ beam of 9 mm diameter. The black surface is reflective chrome while white surface is transparent. The geometry of RLS adopted for FQPM: Lyot outer pupil diameter (5\%\,undersized), Lyot central obstruction diameter (12\%\,oversized), Spider arms (50\%\, oversized).The outer most diameter of RLS is considered to be 3 x Lyot outer pupil diameter. }
\end{itemize}
Image $(c)$ is the complex amplitude after FQPM clearly showing the square intensity pattern diffracted by central obstruction. Image $(e)$ is what is seen by the clowfs at RLS plane. \\
\begin{figure}[h]
   \centerline{
        \resizebox{0.7\columnwidth}{!}{\includegraphics{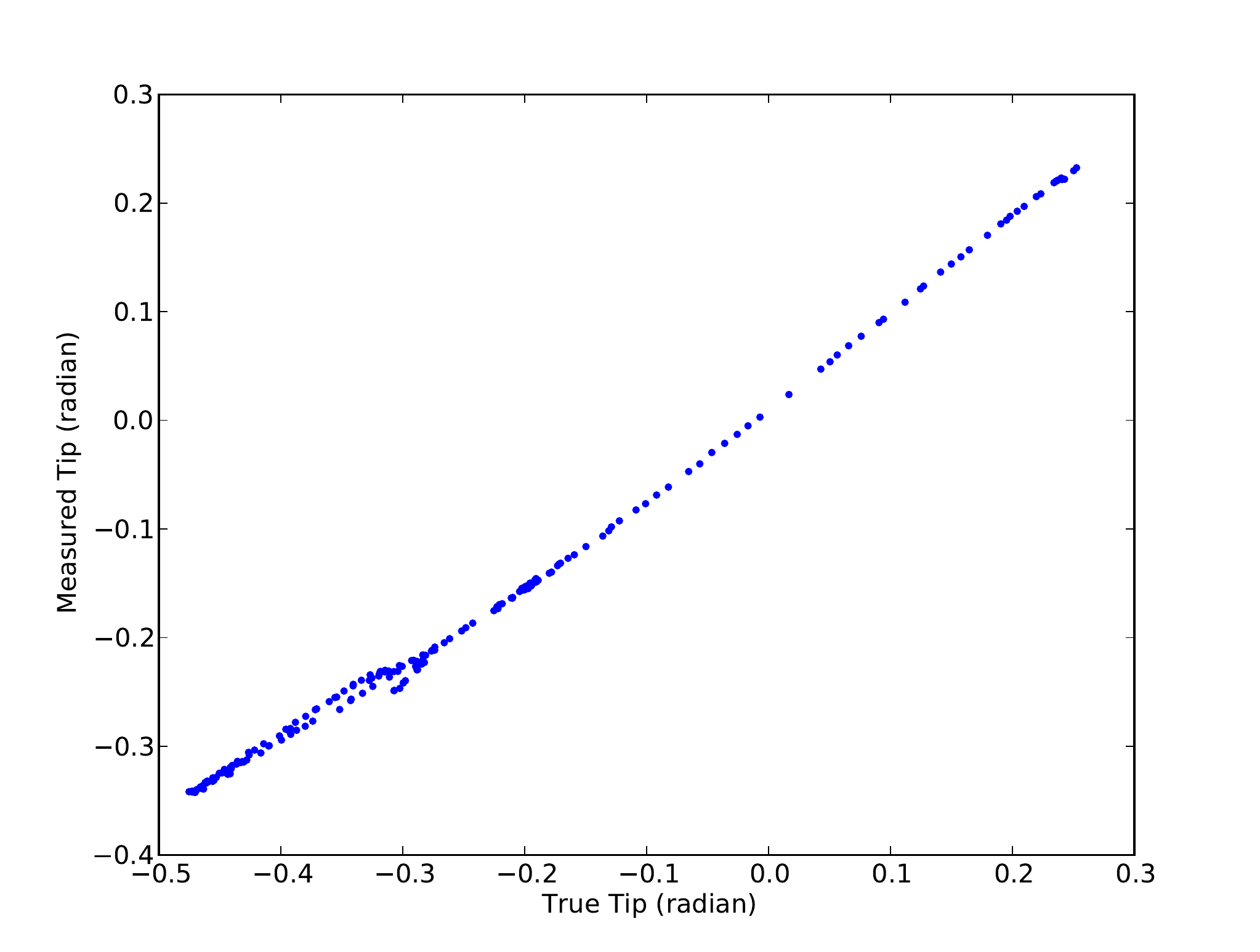}}}
   \caption{Linearity of LPCS response to tip in post-AO188 phase residual.}
  \label{com}
\end{figure}
\begin{figure*}[h]
   \centerline{
        \resizebox{1.2\columnwidth}{!}{\includegraphics{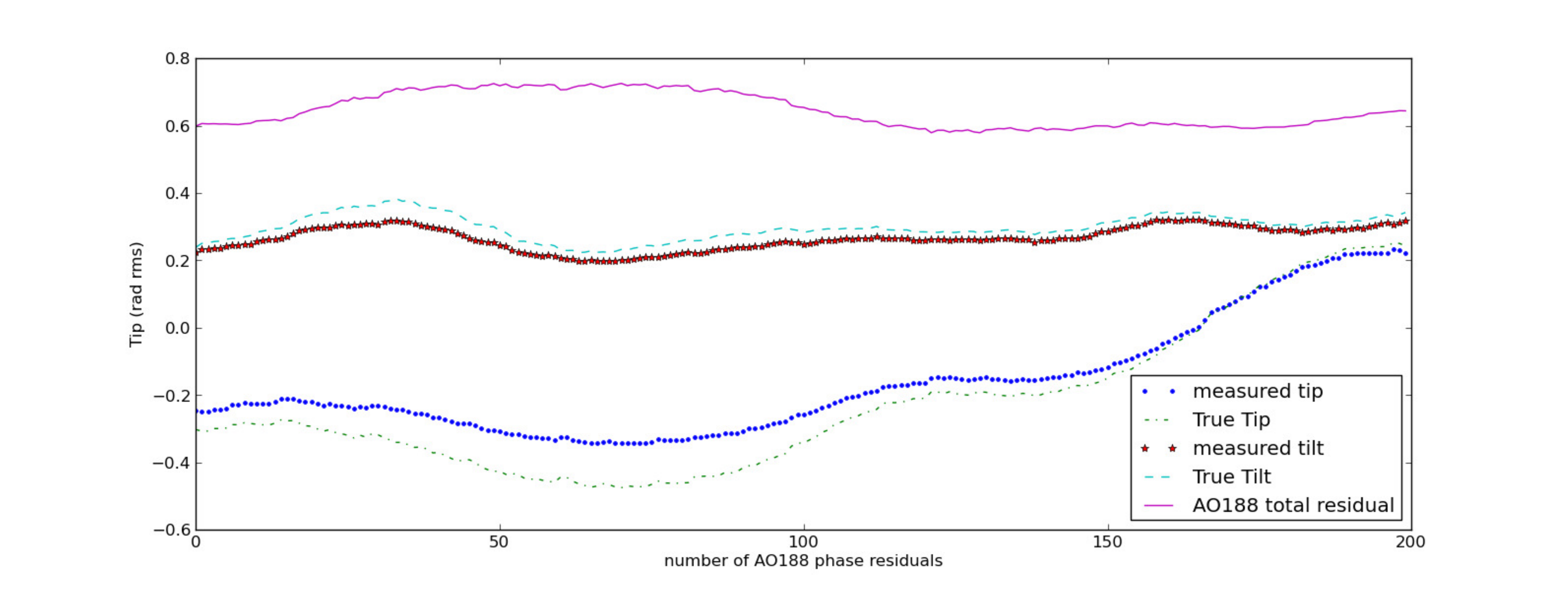}}}
   \caption{
Response of LPCS to tip-tilt in the realistic AO188 phase residual of 200 image series. }
  \label{ao188}
\end{figure*}
\subsection{LPCS Numerical Simulations: Subaru Telescope's AO188-post processed residual phase as wavefront error to LPCS}
\label{ns}
To check the performance of the LPCS, we use the Subaru Telescope's realistic AO188 residual phase series as input to our system. The series consist of 200 residual phase maps with unknown low and high order aberrations ($\approx$ 180 $nm$ phase rms at $\lambda$= 1.6 {\textmu}m). The residual wavefront after encountering the FQPM at the focal plane interfere self-destructively inside the pupil. A wavefront aberration of 0.72 radians is added after FQPM to mimic the quasi-static speckles. The unused diffracted light is then reflected via RLS towards the imaging camera which then measures the tip-tilt that is originally present in the residual phase map. Fig. \ref{com} shows the true tip aberration present in the residual phase map versus tip aberration measured by LPCS showing the linearity range of $\pm$0.1 radians. We also compare the LPCS response to tip-tilt with their actual residual value for 200 phasemaps as shown in Fig. \ref{ao188}. 
The LPCS measurement accuracy for tip-tilt is: \textit{0.05 rad rms} for tip, \textit{0.01 rad rms} for tilt. Fig. \ref{ao188} shows that for small tip-tilt excursions, the LPCS response to low-order aberrations provides a reliable measurement of the tip-tilt. The fidelity of the reconstruction degrades for larger tip-tilt excursion (beyond $\pm$ 0.1 radians), due to non-linearity effects, but the sensor remains well behaved, and in close-loop, would converge toward the reference position, despite the non-linearity. Overall, our system showed the measurement accuracy of 2-12 nm per mode at 1.6 {\textmu}m for FQPM. 
\section{Conclusion}
\label{con}
The lack of accurate pointing control degrades the ability of low inner working angle (IWA) coronagraphs to directly image and characterize exoplanets. To deal with this problematic and delicate issue of pointing control, we have introduced a robust, easy to implement technique to prevent the coronagraphic leaks for low IWA PMCs.
We showed in simulation that with the AO residual phase as an input wavefront, the LPCS is capable of measuring tip, tilt and other low order modes with the accuracy of 2-12 nm at 1.6 {\textmu}m within linear range of $\pm$ 0.1 radians. 
The LPCS measurement is not limited only to tip-tilt as we could also potentially measure defocus and astigmatism as shown in Fig. \ref{cal}. 
Our next step is dedicated to simulate other phase masks such as Roddier \& Roddier (Roddier et al. 1997), Vortex (Mawet et al. 2010), Eight Octant Phase Mask (Murakami et al. 2008a) and the Phase-Induced Amplitude Apodization Complex Phase Mask Coronagraph (Guyon et al. 2010) to verify performance and contrast sensitivity of the LPCS. 

Regarding ELTs and space-based telescopes, the combination of PMC + LPCS promises to provide the sub mili-arcsecond level pointing stability at small IWA making it realistic to directly image the reflected light habitable planets. LPCS is indeed an appealing solution for not only ground-based but for future space missions as well.


\begin{thebibliography}{99}
\bibitem{Ref-Belikov}
Belikov, R., Kasdin, N.~J., \& Vanderbei, R.~J., ApJ, 652 (2006), 833
Guyon, O., A\&A, 404 (2003), 379
\bibitem{Ref-G2003}
Guyon, O., ApJ, 629 (2005), 592
\bibitem{Ref-G2005}
Guyon, O., Pluzhnik, E.~A., Kuchner, M.~J., Collins, B., \& Ridgway, S.~T., ApJ, 167 (2006), 81
\bibitem{Ref-Guyon2009}
Guyon, O., Matsuo, T., \& Angel, R. ApJ, 693 (2009), 75
\bibitem{Ref-Guyon2010}
Guyon, O., Martinache, F., Belikov, R., \& Soummer, R., ApJS, 190 (2010), 220
\bibitem{Ref-Kern}
Kern, B., Guyon, O., Kuhnert, A., Niessner, A., Martinache, F., Balasubramanian, K., Proc. of SPIE, 8864 (2013), 88640R 
\bibitem{Ref-Lloyd}
Lloyd, J.~P., \& Sivaramakrishnan, A., ApJ, 621 (2005), 1153
\bibitem{Ref-Lag}
Lagrange, A., Gratadour, D., Chauvin, G., et~al., A\&A, 493 (2009), L21
\bibitem{Ref-Marois}
Marois, C., Macintosh, B., Barman, T., et~al., Science, 322 (2008), 1348
\bibitem{Ref-Mura}
Murakami, N., Uemura, R., Baba, N., Nishikawa, J., Tamura, M., Hashimoto, N., \& Abe, L., PASP, 120 (2008a), 1112
\bibitem{Ref-Mawet}   
Mawet, D., Serabyn, E., Liewer, K., et~al., ApJ, 709 (2010), 53
\bibitem{Ref-Nem}
Jovanovic, N., Guyon, O., Martinache, Clergeon, C., Singh, G., Vievard, S., Kudo, T., Garrel, V., Norris, B., Tuthill, P., Stewart, P., Huby, E., Perrin, G., Lacour, S., Proc. of AO4ELTs3, Paper 13396 (2013) 
\bibitem{Ref-Roddier}  
Roddier, F., Roddier, C., PASP, 109 (1997), 815
\bibitem{Ref-Rouan}
Rouan, D., Riaud, P., Boccaletti, A., Clenet, Y., \& Labeyrie, A., PASP, 112 (2000), 1479
\bibitem{Ref-Shaklan}     
Shaklan, S.~B., \& Green, J.~J., ApJ, 628 (2005), 474
\bibitem{Ref-Siva}  
Sivaramakrishnan, A., Soummer, R., Sivaramakrishnan, A.~V., Lloyd, J.~P., Oppenheimer, B.~R., \& Makidson, R.~B., ApJ, 634 (2005),1416
\bibitem{Ref-Vogt}
Vogt, F.~P.~A., Martinache, F., Guyon, O., et~al., PASP, 123 (2011), 1434

\end{thebibliography}

\end{document}